# Forecasting of Events by Tweet Data Mining

Bohdan Pavlyshenko

*Ivan Franko Lviv National University, Ukraine, b.pavlyshenko@gmail.com*

**Summary**

This paper describes the analysis of quantitative characteristics of frequent sets and association rules in the posts of Twitter microblogs related to different event discussions. For the analysis, we used a theory of frequent sets, association rules and a theory of formal concept analysis. We revealed the frequent sets and association rules which characterize the semantic relations between the concepts of analyzed subjects. The support of some frequent sets reaches its global maximum before the expected event but with some time delay. Such frequent sets may be considered as predictive markers that characterize the significance of expected events for blogosphere users. We showed that the time dynamics of confidence in some revealed association rules can also have predictive characteristics. Exceeding a certain threshold may be a signal for corresponding reaction in the society within the time interval between the maximum and the probable coming of an event. In this paper, we considered two types of events: the Olympic tennis tournament final in London, 2012 and the prediction of Eurovision 2013 winner.

**Introduction**

The system of microblogs Twitter is a very popular means of users' interaction via short messages (up to 140 characters). The typical feature of Twitter messages is a high density of contextually meaningful keywords. That is why we may assume that while studying microblogs using data mining, we may detect certain semantic relationships between the main concepts and discussion subjects in microblogs. Very promising is to analyze the predictive ability of time dependences of key quantitative characteristics of thematic concepts in Twitter microblogs messages.

The peculiarities of social networks and users' behavior are researched in many studies. In the papers[1,2], the microblogging phenomena were investigated. In [3], M. E. Newman and J. Park showed that social networks differ structurally from other types of networks. Users' influence in Twitter was studied in [4]. Users' behavior in social networks is analyzed in [5]. The paper [6] analyzes the methods of opinion mining of twitter corpus. Several papers are devoted to the analysis of possible event forecasting by analyzing messages in microblogs. In [7], it was studied whether public mood, measured from large-scale collection of tweets posted on Twitter.com, can be correlated or predictive for stock markets. In [8], it is shown that a simple model built from the rate, at which tweets are created about particular topics, can outperform market-based predictors. The paper [9] analyzes films sales based on the discussions in microblogs. In [10], the Twitter activity during media events was studied. The paper [11] studied the phenomenon of resonance in blogospheres which might be caused by real events. The analysis of search engines queries is also used in forecasting [12, 13].

As the examples we take two types of events: the Olympic tennis tournament final in London, 2012 and the prediction of Eurovision 2013 winner.

The distinctive feature of sporting event forecasting is in the fact that tweets reflect bloggers' expectations. These expectations do not always correlate with actual results as the results are often influenced by random factors, the state of sportsmen's training, etc.

On May 18, in the Swedish city of Malmo, the Eurovision Song Contest 2013 took place. The prediction of voting results for determining a winner is very interesting because the participants of the contest were being widely discussed in social media, particularly in Twitter. Moreover, we can assume that active Twitter users, who participated in the discussion, would also vote actively in correspondence with the preferences described in their microblogs. We can therefore expect that the semantic structure of the tweets containing the discussion of the contest will be displayed in the voting results.

In this paper, we construct a set-theoretic model of key tags for Twitter messages and consider the possibility of applying the theories of frequent sets, association rules, and formal concept analysis to event forecasting using tweet minings. We also study the time dynamics of found frequent sets and association rules.

**Theoretical Model**

Let us consider a model, which describes microblog messages. We chose some set of keywords that specify the subjects of messages and these keywords are present in all messages. Then we define a set of microblog messages for our analysis:

$$TW^{kw} = \{tw^{(kw)}_i \mid kw_j \in tw_i, \ kw_j \in Keywords\}. \tag{1}$$

Our next step is to study the basic elements of the theory of frequent sets. Each tweet will be considered as a basket of key terms

$$tw_i = \{w^{tw}_{ij}\}. \tag{2}$$

Such a set is called a transaction. We label some set of terms as

$$F = \{w_j\}. \tag{3}$$

The set of tweets that includes the set $F$ looks like

$$TW^{kw}_F = \{tw_r \mid F \in tw_r; r = 1,...m\}. \tag{4}$$

The ratio of number of transactions that include the set $F$ to the total number of transactions is called a support of $F$ basket and is denoted by $Supp(F)$:

$$Supp(F) = \frac{|TW^{kw}_F|}{|TW^{tw}|}. \tag{5}$$

A set is called frequent if its support value is more than the minimum support, specified by a user

$$Supp(F) > Supp_{min}. \tag{6}$$

Given the condition (6), we find the set of frequent sets

$$L = \{ F_j \mid Supp(F_j) > Supp_{min} \}. \tag{7}$$

For identifying frequent sets, an Apriori algorithm [14, 15] is mainly used. It is based on the principle that the support of a frequent set does not exceed the support of any of its subsets. Based on the frequent sets, we can build the association rules which are considered as

$$X \rightarrow Y, \tag{8}$$

where $X$ is an *antecedent* and $Y$ is a *consequent*. The objects of *antecedent* and *consequent* are the subsets of the frequent set $F$ of considered keywords

$$X \cup Y = F. \tag{9}$$

While finding association rules two major phases are distinguished: the search of all frequent sets of objects and the generation of association rules based on detected frequent sets. Using a frequent set, one can build a large number of association rules, which will be defined by different combinations of features. For the evaluation and selection of useful rules we introduce a number of quantitative characteristics, in particular *support* and *confidence*. *The support* of an association rule shows what part of transactions contains this rule. Since the rule is based on the frequent set of considered keywords, the rule $X \to Y$ has the same support as the set $F$: $X \in F, Y \in F$. Different rules based on the same set have the same support values. The support is calculated by the formula (5). *The confidence* of an association rule shows the probability of the fact that the presence of X attribute subset in the transaction implies the presence of Y attribute subset. Confidence is defined as the ratio of transactions containing the X and Y attribute subsets to the number of transactions containing the X attribute subset only:

$$Conf_{X \to Y} = \frac{\left|TW_{X \cup Y}^{kw}\right|}{\left|TW_X^{kw}\right|} = \frac{Supp_{X \cup Y}}{Supp_X}. \tag{10}$$

An important feature is that different association rules of one and the same set will have different confidence.

Using the theory of formal concept analysis [16,17,18], we consider a formal context as a triple

$$K^{tw(kw)} = \left(TW_s^{(kw)}, Keywords, I_s\right) \tag{11}$$

where $I_s$ is the relation $I_s \subseteq TW_s^{(kw)} \times Keywords$ which describes the connections between messages with the keywords in these very messages. We consider that $(tw_i^{(kw)}, keyword_j) \in I_s$ if a term $keyword_j$ occurs in the message $tw_i^{(kw)}$.

Let us introduce a semantic concept lattice. For some $Ext \subseteq TW_s^{(kw)}$, $Int \subseteq Keywords$, we define the following mappings

$$\begin{aligned} Ext' &= \left\{keyword_j \in Keywords | tw_i^{(kw)} \in Ext : (tw_i^{(kw)}, keyword_j) \in I_s\right\} \\ Int' &= \left\{tw_i^{(kw)} \in TW_s^{(kw)} | keyword_j \in Int : (tw_i^{(kw)}, keyword_j) \in I_s\right\} \end{aligned} \tag{12}$$

The transforms (12) are called Galois transforms. The set $Ext'$ describes the key terms that are peculiar to the documents of $Ext$ set, and the set $Int'$ describes the messages that contain the key terms of $Int$ set.

Now we introduce a semantic concept as a pair

$$Concept = (Ext, Int), \tag{13}$$

which includes the messages from the $Ext \subseteq TW_s^{(kw)}$ set and the key terms from the $Int \subseteq Keywords$ set with the following conditions

$$\begin{cases} Ext' = Int, \\ Int' = Ext. \end{cases} \tag{14}$$

The set $Ext$ is called an extent, and the set $Int$ is called an intent of the semantic concept $Concept$. Have a look at the model of formal concept lattice of Twitter messages, which is based on Galois transformations. The semantic concept lattice can be presented using a Hasse diagram. The volume of the semantic concept is equal to the number of tweets with common keywords which form the intent of this concept. The intent of the concept can be regarded as a frequent set of the words, the support of which is equal to the support of extent of this concept.

Our next step is to consider the notions of *order ideal* and *order filter* for some partially ordered set $(P, \leq)$. An *order ideal* is a subset $J \subseteq P$, for which

$$\forall x \in J, y \leq x \Rightarrow y \in J. \tag{15}$$

An *order filter* is a subset $F \subseteq P$, for which

$$\forall x \in F, y \geq x \Rightarrow y \in F. \tag{16}$$

The use of order ideal and order filter concepts can be effective while analysing the semantic concepts lattice. An order ideal of a concept are the concepts related to it on the Hasse diagram. They are placed below including the concept that fits the lattice infimum. An order filter of a concept is a set of concepts related to it, they are placed above it in the lattice including the concept that fits the lattice supremum.

Similarly to (4) - (9), we can derive the association rules in the semantic concept lattice. These rules reflect the semantic structural relations between keywords. The association rule of some context $K^{tw(kw)} = \left( TW_S^{(kw)}, Keywords, I_S \right)$ is an expression

$$A \to B, \quad A, B \subseteq Keywords \tag{17}$$

The subset $A$ is called an antecedent or a left hand side (LHS) and the subset $B$ is called a consequent or a right hand side (RHS) of the association rule $A \to B$. The important characteristics of association rules are the support $Supp_{A \to B}$ and the confidence $Conf_{A \to B}$, we calculate them by the following formulas:

$$Supp_{A \to B} = \frac{|(A \cup B)'|}{|TW_S^{(kw)}|} \tag{18}$$

$$Conf_{A \to B} = \frac{|(A \cup B)'|}{|A'|} \tag{19}$$

Let us consider the applying of formal concept analysis, frequent sets, and the theory of association rules to two types of events: the Olympic tennis tournament final in London, 2012 and the prediction of Eurovision 2013 winner.

**The analysis of Olympic tennis tournament finals in London, 2012**

For our analysis, we downloaded the messages of Twitter microblogs concerning Summer Olympics in London, 2012. The loading lasted from July 26 to August 15, 2012. We were loading the tweets containing the tags «olympics», «#london2012», «#london2012 tennis», etc. The tweets were loaded into separate files for each selected set of tags. During the analysis time, about 1GB of tweets was downloaded.

Here is the sequence of our analysis. For obvious results that are easy to check, we choose a narrow set of concepts that describe the Olympic tennis tournament final. Then we construct a frequency dictionary of the tweet file with the tags «#london2012». Only the words occurring at least 10 times were taken into consideration. Having removed high-frequency stop words and rarely used terms, we obtained a filtered array of messages. As we know, the final of women's singles tennis tournament was held on August 4, 2012 with Sharapova and Williams playing. And the final of men's singles was held on August 5, 2012 with Murrey and Federer playing.

Our next step is to formalize the semantic frame of our analysis, i.e. to include the words that mean a tournament lap (final), gender (men, women), type (single, double), date (Aug 4, Aug 5), result (gold, silver) athletes' names (Federer , Myrray, Sharapova, Williams). We filter out the set of messages so that they contain only the words from given thematic fields. To construct the Hasse diagram which maps the semantic concept lattice, we used the Lattice Miner software. We created the semantic concept lattice of the frame which was set by the constructed thematic field 1000 thematic concepts were formed, so, we additionally divided the set of semantic features into several subsets. Figure 1 shows the semantic concept lattice representing such concepts as possible tournament dates, sex, the names of contestants for the tweets array with the keywords { #london2012, tennis}. Obviously, in general case it would be preferable to take into account all possible contestants, all dates and all kinds of sports. But it would be difficult to display such a lattice graphically. Therefore, for better clarity we considered a minimum number of concepts.

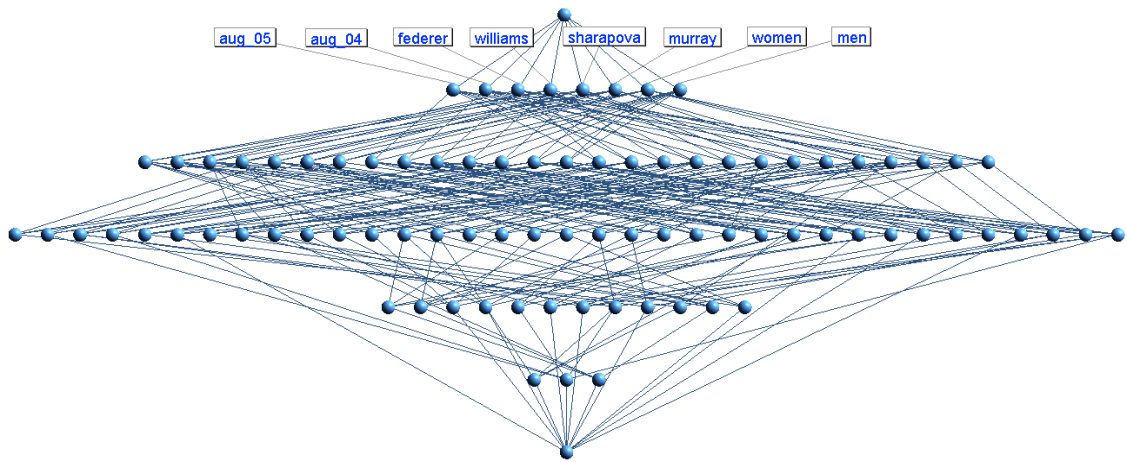

*Fig.1. The semantic concept lattice in tweet array*

Figure 2 shows the semantic concept lattice, where we singled out the order ideal and order filter for the concept (aug_5, federer, murrey, man), its extent is equal to 2% and this is the largest extent for this lattice level. On the basis of this result we can find out that men's final tennis tournament was held on August 5, 2012.

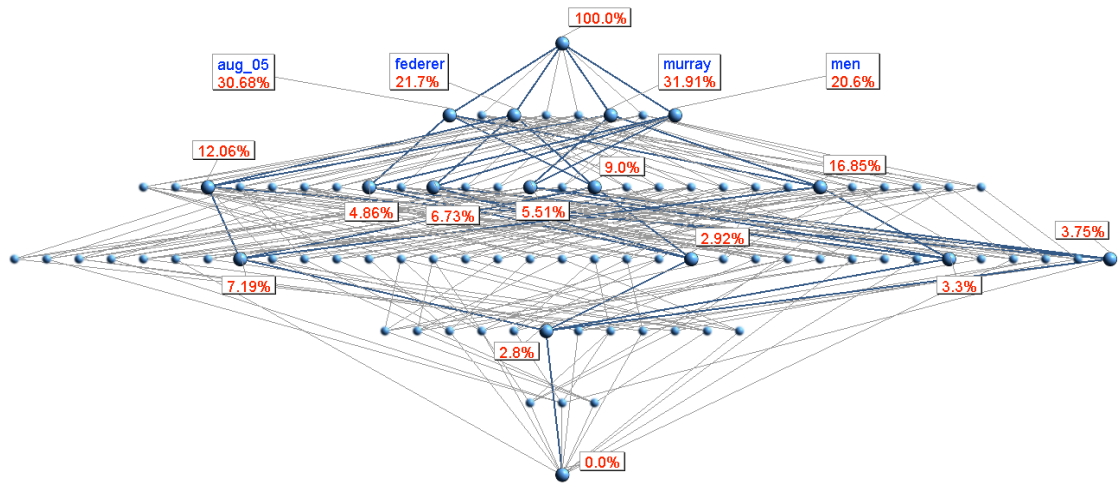

Fig.2 The semantic concept lattice with singled-out order ideal and order filter for the concept {aug_5, federer, murrey, man}

It is evident that in order to choose the concept that corresponds to real event, it is necessary to analyze all possible concepts and to choose the concepts with maximum extent. The concepts with maximum value of extent will be the most likely in reality. Let us consider in details the conceptual relations based on the lattice with the following semantic field

{sharapova, williams, aug_05, aug_04, aug_01, final, wins, gold}.

The results of our analysis are the following values of the extents for analyzed concepts

Ext(sharapova, aug_04, gold)= 3.0 %,

Ext(sharapova, aug_05, gold)= 0.07 %,

Ext(sharapova, aug_01, wins)= 0.04 %,

Ext(sharapova, aug_04, wins)= 1.81 %,

Ext(williams, aug_04, gold)=3.76 %,

Ext(williams, aug_05, gold)=0.79 %,

Ext(williams, aug_01, wins)=0.05 %,

Ext(williams, aug_04, wins)=1.97 % .

We already know it was Williams who won in the final on August 4, 2012. Fig.3 shows the filter and ideal for the concept {sharapova, aug_04, gold}, and Fig.4 shows the fiter and the ideal for the concept {williams, aug_04, gold}. It follows from the given data that

{sharapova, aug_04, gold}< {williams, aug_04, gold}

This inequality reflects the real results of the competition. The difference between the volumes of these two concepts is insignificant. The difference for calculated volumes of the concept {sharapova, aug_04, wins} and {williams, aug_04, wins} is even smaller. However, for similar concepts, with the date aug_5 the difference is more significant and it reflects the outcome of the final, while the concepts with the dates aug_1, aug_4 reflect bloggers' expectations. Fig. 5 shows the order ideal and order filter for the concept {sharapova, aug_05, gold} and Fig. 6 shows the order filter and order ideal for the concept {williams, aug_05, gold}

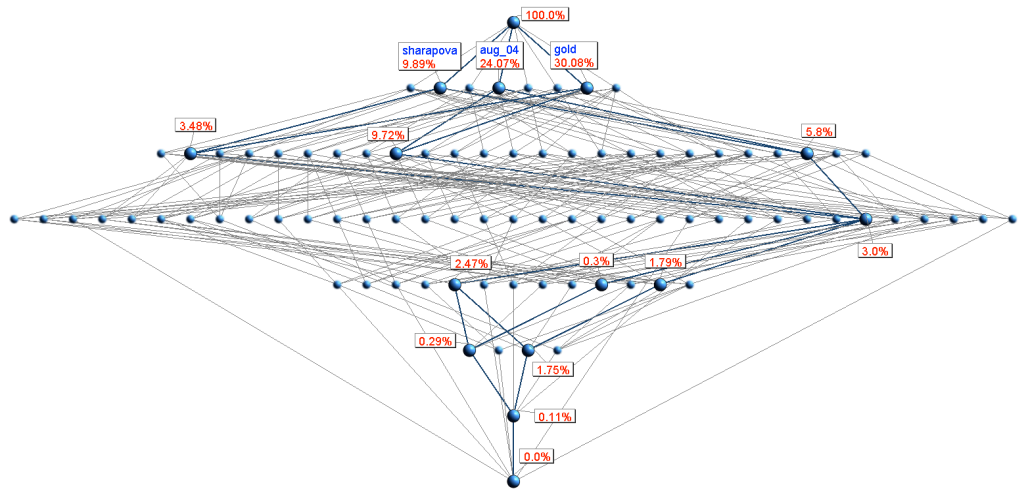

Fig.3. The semantic concept lattice with singled-out order ideal and order filter for the concept {sharapova, aug_04, gold}

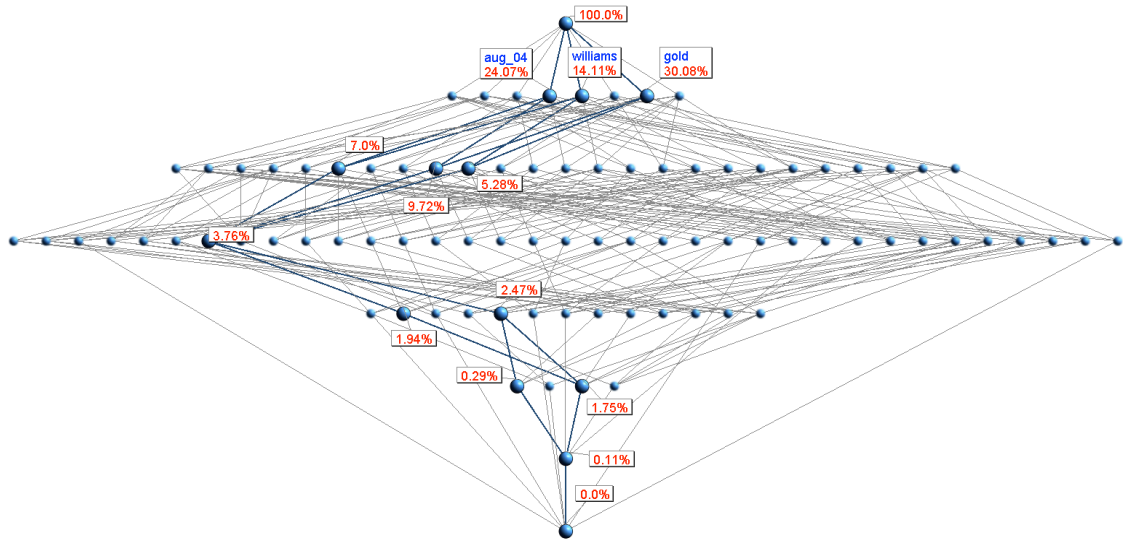

Fig.4 The semantic concept lattice with singled-out order ideal and order filter for the concept {williams, aug_04, gold}

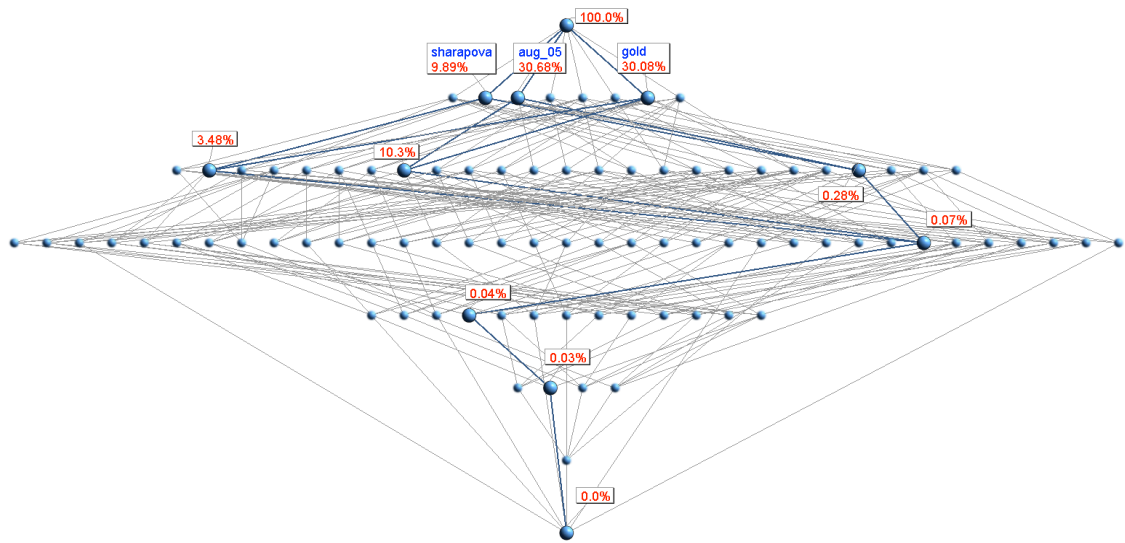

Fig 5 The semantic concept lattice with singled-out order ideal and order filter for the concept {sharapova, aug_05, gold}

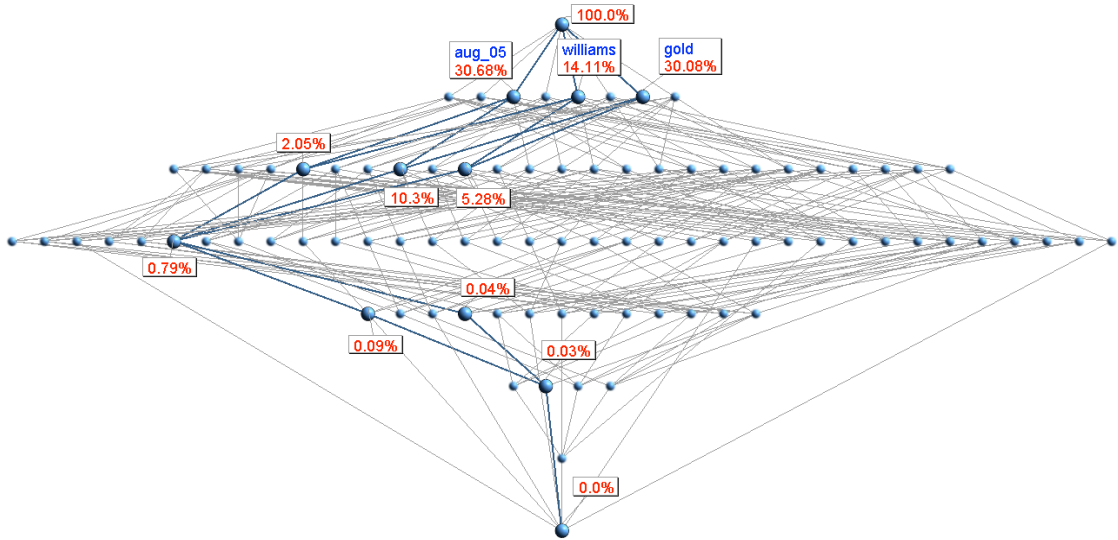

Fig. 6 The semantic concept lattice with singled-out order ideal and order filter for the concept {williams, aug_05, gold}

Let us consider the dynamics of quantitative characteristics of association rules. For our analysis we selected the array with the keywords {#london2012 tennis}. Fig. 7 shows the support dynamics for the association rules Gold-> Sharapova, Gold-> Williams, and the Fig. 8 shows the dynamics of support for these rules. The distinctive features for the obtained curves are the maximums on the very day of the competition and the day after. It means that the support and confidence of association rules on the competition day reflect the expectations and the day after the event - the real result.

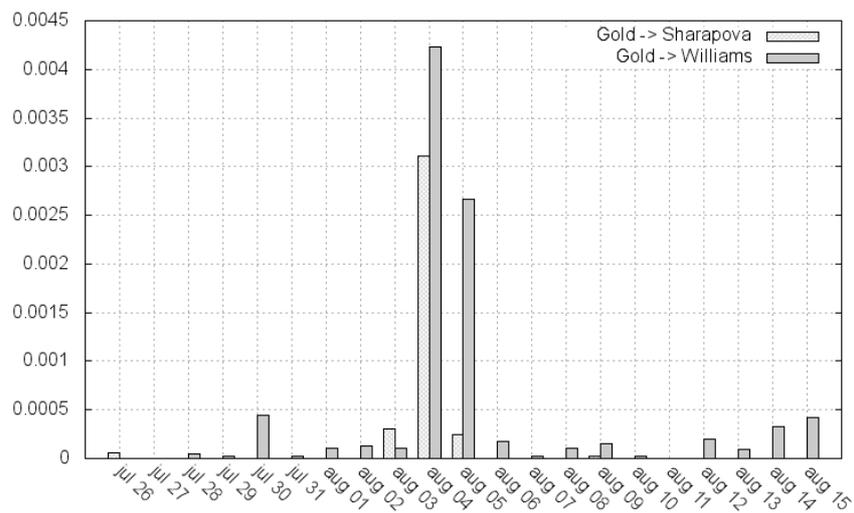

Fig.7 The dynamics of support for the association rules Gold->Sharapova, Gold->Williams.

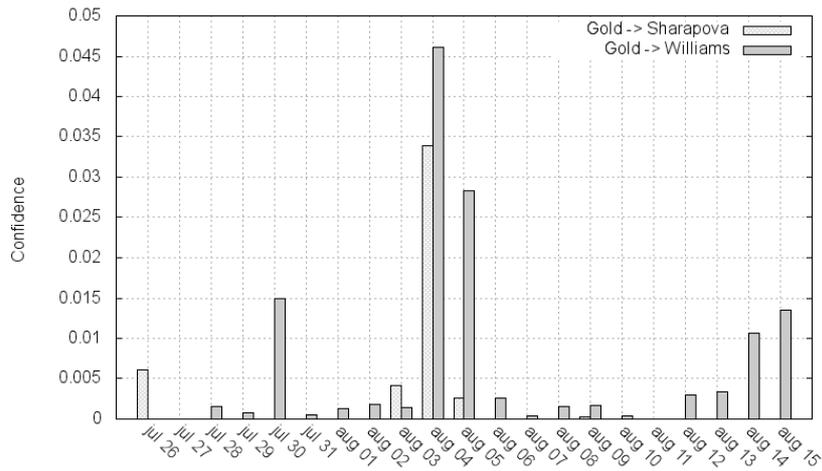

Fig.8 The dynamics of confidence for the association rules
Gold->Sharapova, Gold->Williams.

**Prediction of Eurovision 2013**

On May 17, the day before Eurovision final, we loaded the tweets with the keywords "eurovision win", "eurovision winner". The total amount of tweets was 2400. We conducted the analysis using R language for statistical calculations. The tweets were loaded using the package "twitteR", the analysis of frequent sets and association rules was conducted using the packages "arules", "arulesViz". Then we filtered out the stop words. Our next step was the formation of frequent sets of keywords with the semantic relation to the Eurovision final. Fig.16 shows the graph for the frequent sets with the biggest support.

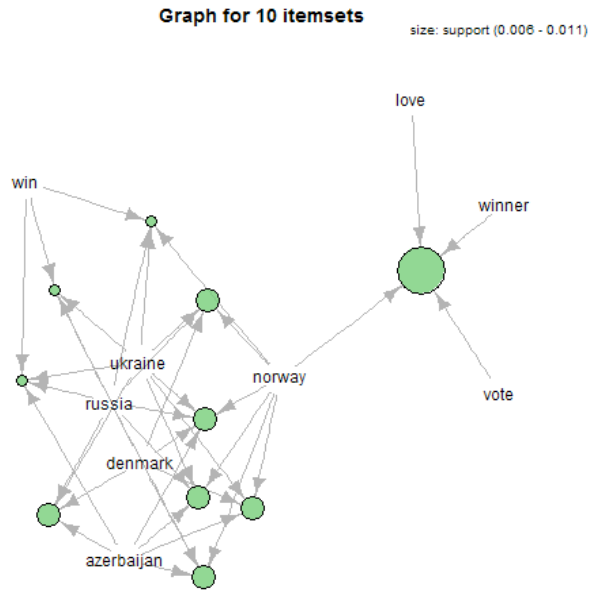

Fig.16 The graph for the frequent sets with the biggest support.

On the basis of selected frequent sets we formed the association rules. The support and confidence of association rules were defined experimentally. We had to form such set of association rules, the right and left parts of which would reflect the semantic concepts relevant to the analyzed event. In particular, these are the country names, such keywords as "win",

"winner", "favorite", "vote", "love", etc. Obtained association rules were sorted out by the size of support. Here are some of the examples of association rules with the biggest value of support on the left side of the association rule. Obtained association rules are shown in the tables 2-4.

*Table 2. The association rules with two keywords on the left side.*

|    | Antecedent | consequent | support | confidence | lift |
|---|---|---|---|---|---|
| 1 | denmark, norway | win | 0.014610390 | 0.9000000 | 1.3137441 |
| 2 | denmark, favourites | win | 0.011363636 | 1.0000000 | 1.4597156 |
| 3 | azerbaijan, norway | win | 0.011363636 | 0.8750000 | 1.2772512 |
| 4 | denmark, ukraine | win | 0.008116883 | 0.8333333 | 1.2164297 |
| 5 | azerbaijan, russia | win | 0.008116883 | 0.8333333 | 1.2164297 |
| 6 | azerbaijan, denmark | win | 0.008116883 | 0.7142857 | 1.0426540 |
| 7 | finland, sweden | win | 0.008116883 | 1.0000000 | 1.4597156 |
| 8 | russia, ukraine | win | 0.006493506 | 0.8000000 | 1.1677725 |
| 9 | azerbaijan, ukraine | win | 0.006493506 | 0.8000000 | 1.1677725 |
| 10 | norway, ukraine | win | 0.006493506 | 0.8000000 | 1.1677725 |

*Table 3. The association rules with three keywords on the left side*

| # | Antecedent | consequent | support | confidence | lift |
|---|---|---|---|---|---|
| 14 | azerbaijan, russia, ukraine | win | 0.006493506 | 0.8000000 | 1.1677725 |
| 15 | norway, russia, ukraine | win | 0.006493506 | 0.8000000 | 1.1677725 |
| 16 | denmark, russia, ukraine | win | 0.006493506 | 0.8000000 | 1.1677725 |
| 17 | azerbaijan, norway, ukraine | win | 0.006493506 | 0.8000000 | 1.1677725 |
| 18 | azerbaijan, denmark, ukraine | win | 0.006493506 | 0.8000000 | 1.1677725 |
| 19 | denmark, norway, ukraine | win | 0.006493506 | 0.8000000 | 1.1677725 |
| 20 | azerbaijan, norway, russia | win | 0.006493506 | 0.8000000 | 1.1677725 |
| 21 | azerbaijan, denmark, russia | win | 0.006493506 | 0.8000000 | 1.1677725 |
| 22 | denmark, norway, russia | win | 0.006493506 | 0.8000000 | 1.1677725 |
| 23 | azerbaijan, denmark, norway | win | 0.006493506 | 0.8000000 | 1.1677725 |

*Table 4. The association rules with four keywords on the left side*

| # | Antecedent | consequent | support | confidence | lift |
|---|---|---|---|---|---|
| 24 | azerbaijan, norway, russia, ukraine | win | 0.006493506 | 0.8000000 | 1.1677725 |
| 25 | azerbaijan, denmark, russia, ukraine | win | 0.006493506 | 0.8000000 | 1.1677725 |
| 26 | denmark, norway, russia, ukraine | win | 0.006493506 | 0.8000000 | 1.1677725 |
| 27 | azerbaijan, denmark, norway, ukraine | win | 0.006493506 | 0.8000000 | 1.1677725 |
| 28 | azerbaijan, denmark, norway, russia | win | 0.006493506 | 0.8000000 | 1.1677725 |
| 29 | azerbaijan, denmark, norway, russia, ukraine | win | 0.006493506 | 0.8000000 | 1.1677725 |

The next step of our study is the graphic representation of association rules. To build a graphical display of association rules, we use a VizRules package for R language. Fig. 17 shows the association rules with the highest support values. Fig.18 shows the formation of association rules based on keywords. Fig.19 shows the graphic presentation of the matrix of grouped association rules.

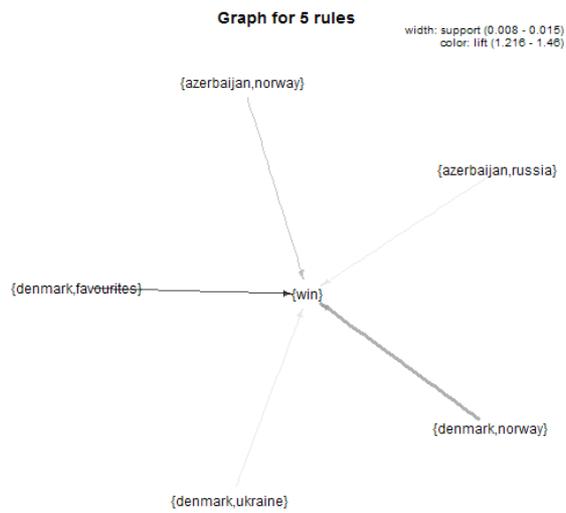

Fig 17. Association rules with the biggest support values.

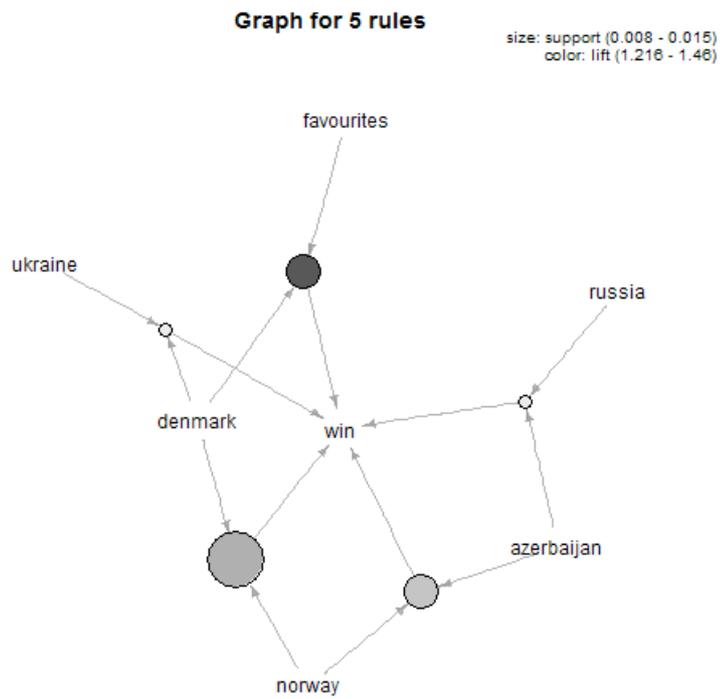

Fig.18 The formation of association rules on the basis of keywords.

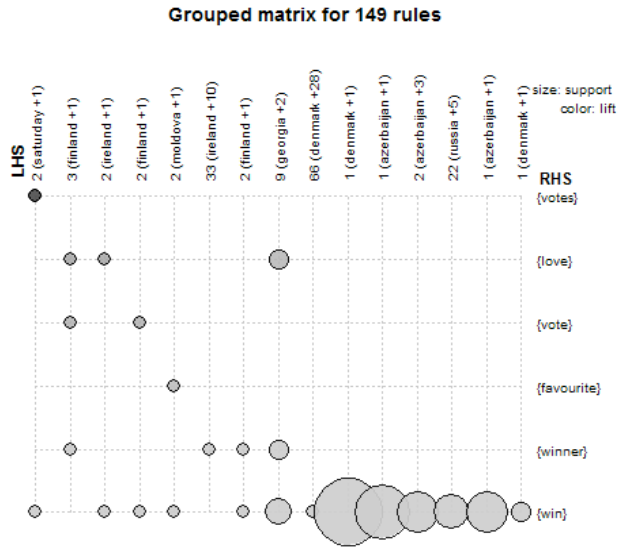

Fig. 19 The matrix of grouped association rules.

On the basis of obtained results one can draw a conclusion that Denmark was the leader among all favourites, the next three places go to Ukraine, Russia and Ireland. The anounced results of the final were: 1st place - Denmark, 2nd place - Azerbaijan, 3rd place - Ukraine, 4th place - Norway, 5th place - Russia. As follows from our results and the final results, our data mining analysis has correctly detected the winner and the real rating of top places .

**Conclusions**

The model of tweets studied in the paper was tested in analyzing and forecasting the events of two types. The first type corresponds to sporting events where users' expectations can be corrected by random factors and real level of athletes' training, which may differ from fans' expectations. Thus, the value of support of corresponding concepts for the finalists were similar before the tournament and significantly different after the tournament, this is what the results of the final represent. Using the formal concept model in the analysis of Twitter microblogs messages enables to detect effectively the semantic relations between such thematic concepts of sporting events as time of a competition, sex of competitors, types of sports, names of competitors, the results of competitions, and names of winners. Along with the mapping of real facts, the semantic concept lattice displays bloggers' forecasts and expectations.

As the example of the second type of events, we considered the final of Eurovision Song Contest 2013. The peculiarity of this event type is caused by the fact that the users who discuss contest participants represent a group of those who participate actively in the telephone poll. The difference between forecasting of this very event type and sporting events is that the influence of random factors is significantly smaller, as the contestants and their songs are known in advance and the blogosphere participants have already formed their own assessment of participants, which is unlikely to change, while the sports result is much more likely to be different from expectations. That is why the second type of events is characterized by high correlation between the results of the event and the results of previous data mining of tweets.